# Block spin transformation on the dual lattice and monopole action

Tsuneo Suzuki [a], Yoshimi Matsubara [b], Shinji Ejiri [a], Kazuya Yamada [a] and Natsuko Arasaki [a]

[a]Department of Physics, Kanazawa University, Kanazawa 920-11, Japan

[b]Nanao Junior College, Nanao, Ishikawa 926, Japan

To find a perfect lattice action in terms of monopole action on the dual lattice, we performed simulations of a monopole effective action obtained numerically from vacuum configurations in SU(2) QCD. Although the Polyakov loop behavior near $T_c$ is well reproduced by the action, a small but repulsive term is needed in addition to get the string tension correctly. It is reported also a monopole effective action in $SU(3)$ QCD which is expressed by one kind of monopole currents.

## 1. Introduction

It has been found that confinement phenomena seem to be well reproduced by abelian link fields alone in the maximally abelian (MA) gauge in $SU(2)$ QCD [1–5]. The abelian dominance suggests the existence of an effective $U(1)$ theory describing confinement. The purpose of this talk is to report our study toward a perfect lattice action in terms of abelian monopole currents on the dual lattice based on our standpoint that monopole condensation is the confinement mechanism in QCD as suggested by 'tHooft[6]

## 2. Monopole action for SU(2) QCD

After the abelian projection, one can separate out abelian link fields $u(s,\mu)$ as

$$U'(s,\mu) = V(s)U(s,\mu)V^{\dagger}(s+\hat{\mu}) \equiv c(s,\mu)u(s,\mu).$$

The abelian dominance in MA gauge means that the set of operators composed of $u(s,\mu)$ alone are enough for explaining the essential features of confinement[4]. Then there must exist an effective abelian action $S_{eff}(u)$ describing confinement. The action is given by

$$S_{eff}(u) = -\ln(\int Dc e^{-S(u,c)}\delta(X^{\pm})\Delta_F(u,c)),$$

where $X^{\pm} = 0$ is the gauge-fixing condition and $\Delta_F(u,c))$ is the Fadeev-Popov determinant. However it was found that $S_{eff}(u)$ is not fixed to be local[4].

Shiba and one of the authors (T.S.)[7] have tried to determine a monopole effective action defined as[8]

$$\begin{aligned}\exp(-S[k]) &= \int Du\delta(k,u)\exp(-S_{eff}(u)),\\ \delta(k,u) &\equiv \delta(k_\mu(s) - \frac{1}{2}\epsilon_{\mu\nu\rho\sigma}\partial_\nu m_{\rho\sigma}(s+\hat{\mu})).\end{aligned}$$

performing a dual transformation numerically. We have also considered $n^3$ extended monopoles defined on a sublattice with the spacing $b = na$[9]. This corresponds to making a block-spin transformation on the dual lattice as seen from

$$\begin{aligned}e^{-S^{(n)}[k^{(n)}]} &= (\prod_{s,\mu}\sum_{k_\mu(s)=-\infty}^{\infty})(\prod_s \delta_{\partial'_\mu k_\mu(s),0})\\ &\quad (\prod_{s,\mu}\delta(k_\mu^{(n)}(s) - F(k_\mu(s)))e^{-S[k]},\end{aligned}$$

$$F(k_\mu(s)) = \sum_{i,j,l=0}^{n-1} k_\mu(ns + (n-1)\hat{\mu} + i\hat{\nu} + j\hat{\rho} + l\hat{\sigma}).$$

The effective monopole actions $S^{(n)}[k^{(n)}]$ for $n = 1 \sim 4$ have been fixed successfully from the ensemble $\{k_\mu^{(n)}(s)\}$ calculated from vacuum configurations on $24^4$ lattice by extending the Swendsen method[10]. The monopole action adopted is composed of various two-point current-current interactions $S[k] = \sum_i f_i S_i[k]$, the first of which is the self-coupling term $S_1[k] = \sum k_\mu^2(s)$.

The summary of the monopole action determined in $SU(2)$ is the following[7]:

1. A compact and local form of the monopole action is obtained.

2. $f_i$ look volume independent.($8^4 \sim 24^4$)

3. Monopole condensation is seen to occur for smaller $\beta$ from energy-entropy balance.

4. $f_i$ looks to depend only on $b = na$, not on the extendedness nor $\beta$. There is a kind of scaling.

5. If the scaling of $S(k_\mu^{(n)}(s))$ remains true even on the $\infty$ lattice, the SU(2) QCD vacuum is always (for all $\beta$) in the monopole condensed and then color confined phase.

6. The SU(2) monopole action seems on or near to the renormalized trajectory of the block spin transformation as seen from the renormalization flow in Fig. 1.

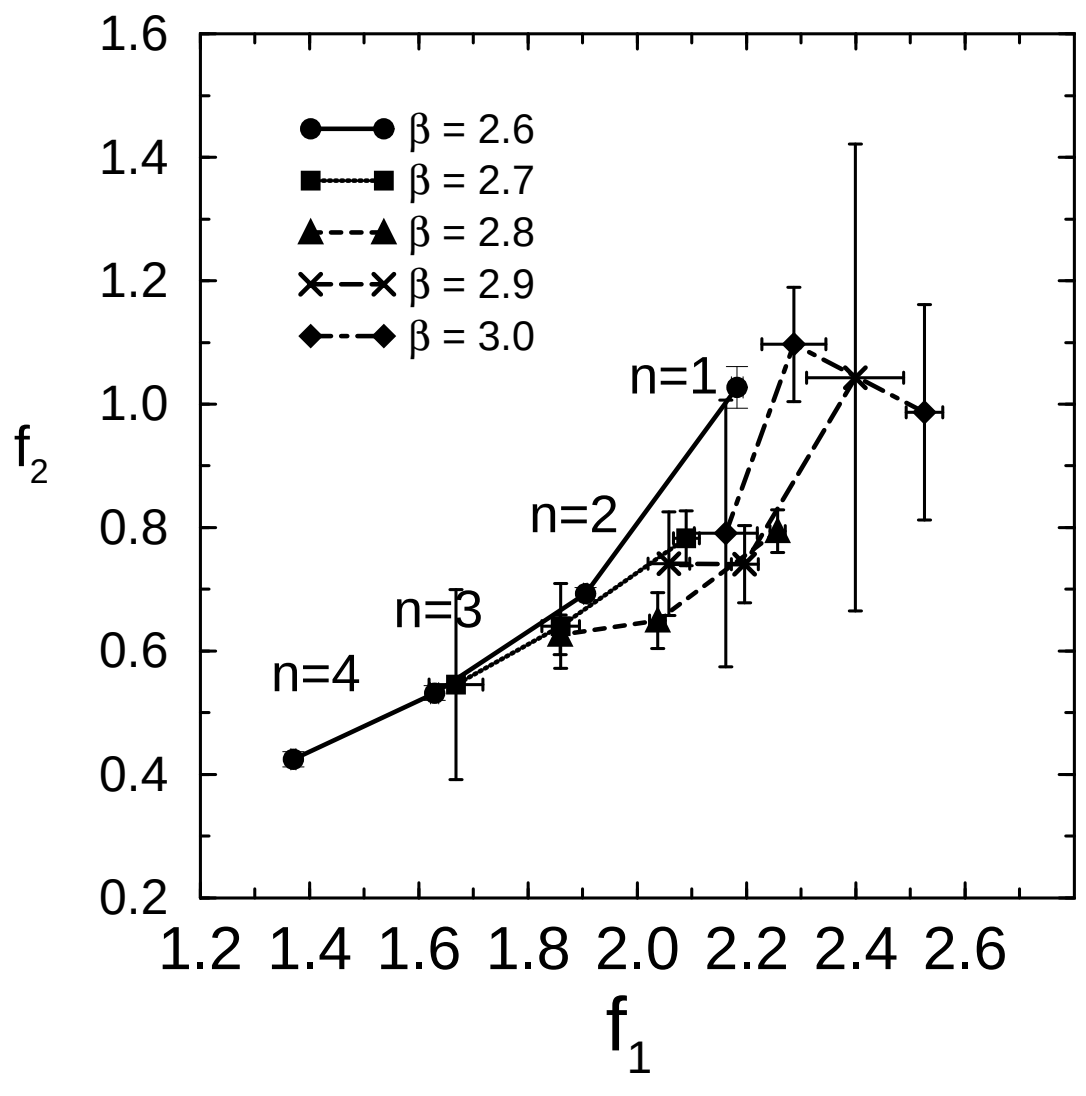

Figure 1. The $f_1 - f_2$ cross section of the renormalization flow of the block-spin transformation on the dual lattice.

## 3. Simulations of monopole action and the renormalized trajectory

The renormalized trajectory must exist near the action obtained above. Let us search for the trajectory, performing simulations of the monopole action. First we adopt the following action:

| $b$ | $f_1$ | $f_2$ | $f_3$ | $f_4$ | $f_5$ | $f_6$ |
|---|---|---|---|---|---|---|
| 5.55 | 1.91 | 0.69 | 0.42 | 0.17 | 0.11 | 0.09 |
| 7.15 | 1.74 | 0.58 | 0.40 | 0.14 | 0.11 | 0.10 |
| 9.20 | 1.52 | 0.49 | 0.34 | 0.12 | 0.10 | 0.08 |

Table 1
Coupling constants for typical $b(\times 10^{-3}\Lambda_L^{-1})$

$$S(k) = \sum k_\mu(s) D(s,s') k_\mu(s')$$

$$D(s,s') = f_1\delta_{s,s'} + f_2\delta_{s+\mu,s'} + f_3\delta_{s+\nu,s'}$$
$$+ f_4\delta_{s+\mu+\nu,s'} + f_5\delta_{s+\nu+\rho,s'} + f_6\delta_{s+2\mu,s'},$$

where the coupling constants are chosen to take the (approximate) values determined in [7] and are shown in Table 1. Our method of the simulation is the following:

1. Cold start (all $k_\mu(s) = 0$) or hot start (random number).

2. Consider a plaquette $(s,\mu,\nu)$ on the dual lattice.

3. Generate a random number $K$ taking $\pm 1$.

4. Change the monopole currents on the plaquette $(s,\mu,\nu)$ as

   $$k_\mu(s) + K, \quad k_\nu(s+\hat\mu) + K,$$
   $$k_\mu(s+\hat\nu) - K, \quad k_\nu(s) - K$$

   This does not violate the current conservation.

5. Make the Metropolis selection.

6. Evaluate abelian Wilson loops, the string tension and the length of monopole loops.

The method works well and we have obtained the following results:

1. The thermalization is obtained rapidly except near the transition temperature.
2. The action with quadratic interactions alone can reproduce the Polyakov loop behavior. Namely, the Polyakov loop vanishes in the confinement phase, but begins to rise rapidly at the critical temperature where $f_1 \sim 2.0$ as seen from the original monopole action in $T \neq 0$ $SU(2)$ QCD[11]. See Fig. 2.

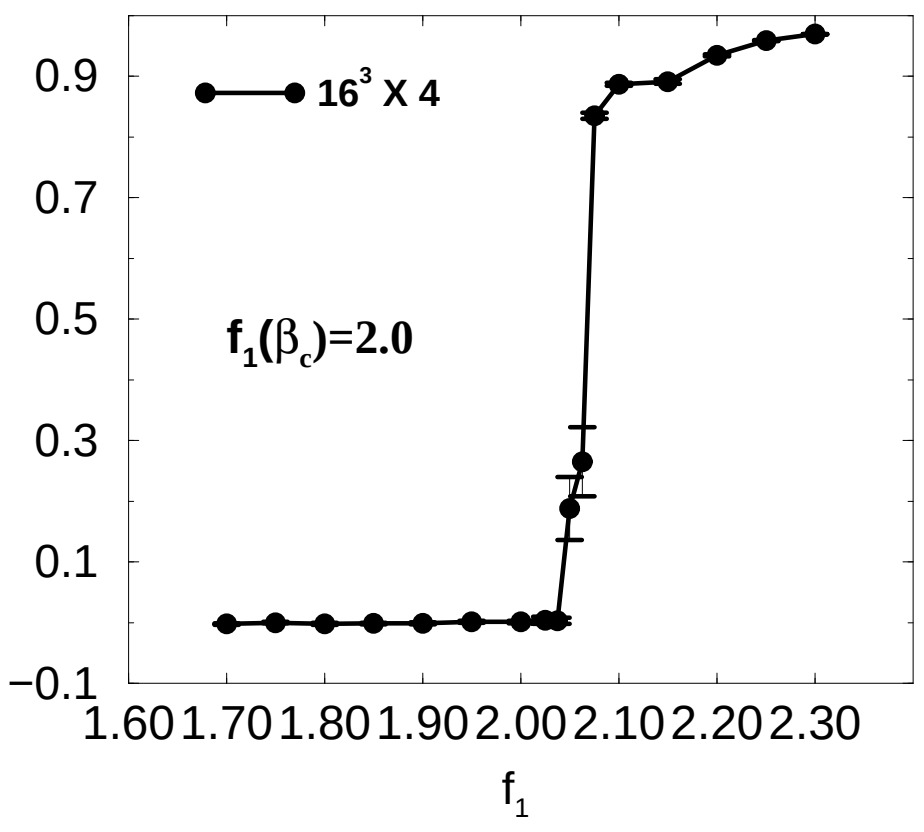

Figure 2. The Polyakov loop behavior from the simulation of the monopole action.

3. It gives, however, a longer monopole loop and a larger string tension than expected. We may need a small term which gives a repulsive force between monopole currents in addition. As an example, we have added the following terms:

   $$f_7 \sum |k_\mu(s)||k_\nu(s)| + f_8 \sum |k_\mu(s)| \times |k_\mu(s+\hat{\mu})| + f_9 \sum |k_\mu(s)||k_\mu(s+\hat{\nu})|$$

   With a small value for $f_7 \sim f_9$, we get a nice fit of the loop length and the string tension on small lattices. Finite-size effects are seen to be small.

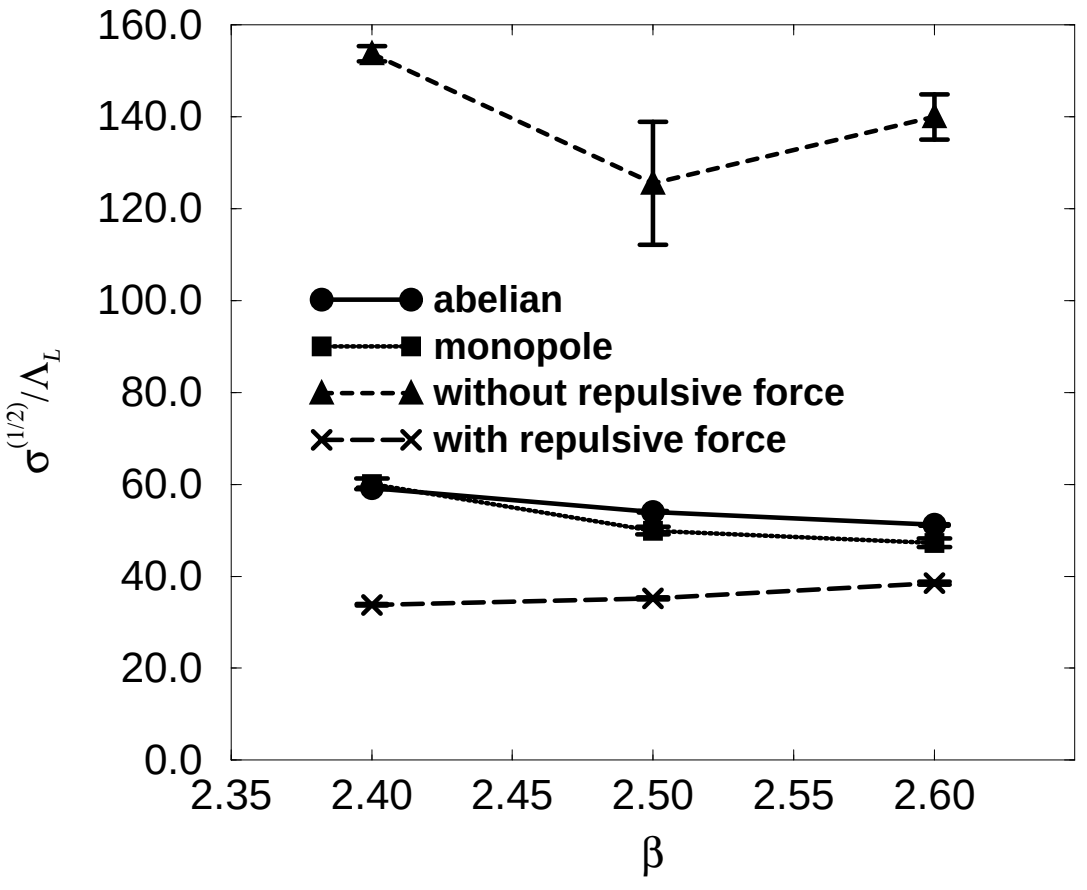

Figure 3. String tension versus lattice volume.

4. The length of the long loop and the value of the string tension are very sensitive to a change of the small parameters $f_7 \sim f_9$. It is however very difficult to determine correctly such small repulsive terms which make the monopole length adequate from the vacuum configurations, since the monopoles in the thermalized vacua have much the same length.

## 4. Monopole action for SU(3) QCD

How about the case of SU(3) QCD? There are two independent (three with one constraint $\sum_{i=1}^{3} k_\mu^i(s) = 0$) currents. When considering the two independent currents, their entropies are difficult to evaluate. Hence let me first try to evaluate the effective monopole action, paying attention to only one monopole current.

The monopole action in $SU(3)$ QCD is obtained for $\beta = 5.0 \sim 6.0$[7]. Lattice sizes considered are $8^4$ (for $T = 0$ system) and $12^3 \times 4$ (for $T \neq 0$ system). Only the smallest monopole is taken into account as the first step.

The results in the $T = 0$ case are the following:

1. Action is fixed in a local form: $f_1 \gg f_2 \sim f_3 > f_4 \sim f_5 \sim f_6$
2. It is very interesting to see that monopole condensation occurs at least for $\beta \leq 5.6$

as seen from energy-entropy balance. Note that the $\ln 7$ line shows the entropy per unit monopole length. This is the first result showing the monopole condensation in $SU(3)$ QCD.

3. The form of the monopole action is very similar to those of the SU(2) QCD and of the Wilson compact QED in the strong coupling region.

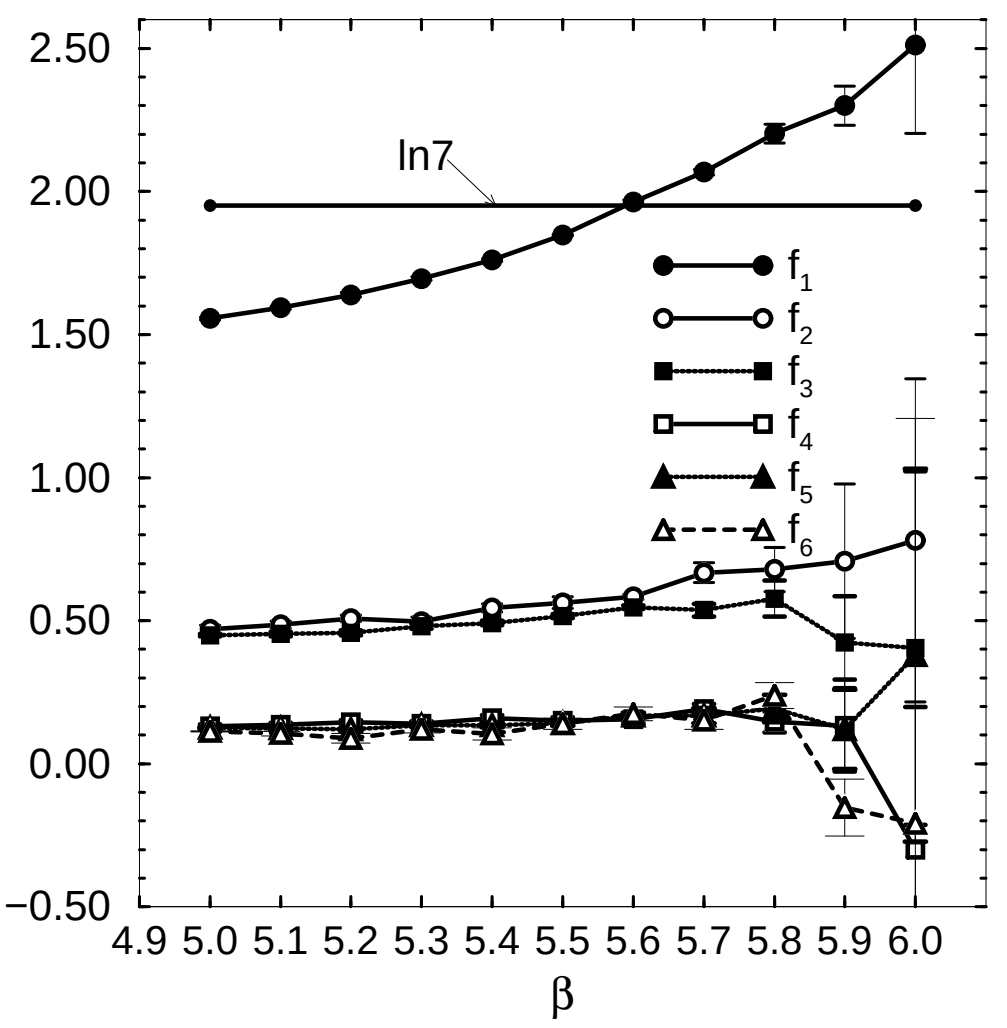

Figure 4. The effective monopole action in $T = 0$ $SU(3)$ QCD.

The action in the $T \neq 0$ case is also fixed. Near the transition temperature, there are some discrepancies between cold and hot start monopole actions which correspond to the first order transition. However, a clear hysteresis curve is not seen because the space extent of our lattice is too short. We are studying a larger lattice to see a clear signal of the first-order transition. To study the mechanism of the first-order transition, we have to analyse the behaviors and dynamics of three kinds of monopoles.

## 5. Final remarks

The followings are interesting subjects to be studied in near future.

1. A block-spin transformation on the dual lattice considering extended monopoles is very interesting as in $SU(2)$ QCD. Is the action a function of $b(= na)$ alone as in SU(2) QCD?

2. In the finite-temperature $SU(3)$ QCD, we have to study especially the interplay of three kinds of monopoles in the role of the first-order transition.

3. To get a monopole action in full $SU(3)$ QCD with dynamical fermions is very important to see the relation between chiral breaking and confinement.

This work is financially supported by JSPS Grant-in Aid for Scientific Research (B) (No.06452028).